\documentclass[authoryear,12pt,3p]{jowarticle}

\usepackage[english]{babel}
\usepackage[utf8]{inputenc}
\usepackage{amsmath}
\usepackage{graphicx}
\usepackage{setspace}
\usepackage[colorinlistoftodos]{todonotes}
\usepackage{natbib}

\makeatletter
\renewcommand\section{\@startsection{section}{1}{\z@}{-3.25ex plus -1ex minus -.2ex}{1.5ex plus .2ex}{\normalsize\bf}}
\renewcommand\subsection{\@startsection{subsection}{2}{\z@}{-3.25ex plus -1ex minus -.2ex}{1.5ex plus .2ex}{\normalsize\bf}}
\renewcommand\subsubsection{\@startsection{subsubsection}{3}{\z@}{-3.25ex plus -1ex minus -.2ex}{1.5ex plus .2ex}{\normalsize\bf}}

\usepackage{enumerate}
\usepackage{graphicx}
\usepackage[round]{natbib}
\usepackage{amssymb,amsmath, amscd, amsthm, mathrsfs}

\usepackage{ bbold }
\theoremstyle{definition}

\newtheorem*{con}{Conjecture}

\theoremstyle{remark}

\numberwithin{equation}{section}

\makeatother

\begin{document}
\begin{frontmatter}
\title{Where Does General Relativity Break Down?}

\author{James Owen Weatherall}\ead{weatherj@uci.edu}
\address{Department of Logic and Philosophy of Science \\ University of California, Irvine}

\date{ }

\begin{abstract}
It is widely accepted by physicists and philosophers of physics alike that there are certain contexts in which general relativity will ``break down''. In such cases, one expects to need some as-yet undiscovered successor theory.  This paper will discuss certain pathologies of general relativity that might be taken to signal that the theory is breaking down, and consider how one might expect a successor theory to do better.  The upshot will be an unconventional interpretation of the ``Strong Cosmic Censorship Hypothesis''.
\end{abstract}
\end{frontmatter}
\doublespacing
\section{Introduction}\label{sec:intro}

As is often observed by physicists and philosophers alike, general relativity exhibits some striking, even pathological, features \citep{EarmanBCWS}.  The best known of these are singularities, such as the Big Bang singularity in expanding cosmological spacetimes or the singularities associated with black holes.  A common refrain, particularly among physicists, is that these singularities are ``unphysical'' in some sense.  That is: even if we take general relativity to be approximately true, we should not infer that truly singular behavior is physically possible, even though the theory appears to predict it.  Instead, we should take singularities to indicate that general relativity will ``break down'' in application to certain situations, in the sense that it will fail to be representationally adequate.  An adequate description of these physical situations would require a new theory, generally expected to be a quantum theory of gravity.  This expectation is often summarized by the claim that quantum gravity will ``resolve'' singularities \citep{Bojowald}.

This paper is concerned with the question posed in the title: Where does general relativity (GR) break down?  The word ``where'' in this question is playing a double role: on the one hand, I will ask ``for what physical situations, putatively within the domain of applicability of GR'' the theory should be understood to be inadequate; and on the other hand, I wish to insist that any satisfactory understanding of the breakdown of GR must be ``local'' in the sense that, at least in some cases, the breakdown can be associated with some region of space and time.  That is, there is somewhere that the breakdown occurs, in the sense that given a relativistic spacetime, one should expect to be able to identify some open set, with compact closure, such that the breakdown is localized to that region and outside of that region GR is representationally adequate.

In the background of this question is a certain ideology, very common in physics, that our current theories are merely ``effective'' descriptions of the world, approximately correct in some regimes, but certain to fail in others.\footnote{This ideology is at odds with how the truth and falsity of scientific theories has traditionally been understood by general philosophers of science, but has been widely accepted among physicists since the 1970s.  For recent work connecting these ideas about ``effective'' theories and scientific realism, see \citep{Williams}.}  That any failures of GR can be localized in the way I have indicated is essential to making sense of how the breakdown of GR is understood in physics.  But in fact, this ideology usually involves a further claim, which is that not only can we be confident that our current theories, including general relativity, will break down in some regimes, but in fact we can \emph{anticipate} where our current theories will fail.  They will fail at high energies.

In more detail: usually one speaks of effective theories in the context of renormalization group methods in particle physics.  In this context, there is a single parameter, the energy of a system, whose value signals whether the theory will break down.  When the energy associated with some system, process, or interaction approaches or exceeds a characteristic value---the ``cutoff''---associated with a given theoretical description, that description should be expected to break down.  This reasoning, applied to GR, suggests that there, too, for sufficiently large energies, the theory will fail.

But there are problems with this line of thought.  One problem is that it is not clear how to think about the high-energy regime in general relativity.  Another problem is that, even if we solve the first problem, it is not clear that all of the ``pathologies'' of GR can be identified with any one regime, high-energy or not.  Addressing these two problems will be the focus on the present paper.

I will proceed as follows.  I will begin by discussing what one might mean by the ``high energy regime'' in general relativity.  I will argue that this concept is not as neatly described as one might hope for, and there is not single scalar quantity that always and only ``becomes large'' in the high-energy regime.  Next, I will consider whether general relativity breaks down only in this high energy regime.  Here I consider two senses in which the theory breaks down: singularities and certain Cauchy horizons.   In the final section, I will discuss how an important conjecture, known as the strong cosmic censorship hypothesis (SSCH), bears on these issues.  The principal claim of this section will be that if a certain version of the SSCH is true, then singularity resolution and ``Cauchy horizon resolution'' are deeply connected.

\section{What is the ``high-energy regime'' in general relativity?}

Before we can evaluate whether general relativity should be expected to break down (only) in the high energy regime, we must first establish what the ``high-energy regime'' is in the theory.  Doing so is subtle.  There is no fully satisfactory answer.

Consider, for instance, an answer motivated by how the high-energy regime is usually characterized in particle physics. There one associates ``high energy'' with large particle/field kinetic energies (or, high temperature, as in the early universe).  The idea, here, is that large kinetic energy leads to high-energy scattering and that general relativity will break down as (part) of the description of those scattering events.  On this proposal, the high-energy regime is signalled by ``large'' stress-energy---with large in scare quotes because stress-energy is a tensor, and it is not clear that there is a physically meaningful and observer independent characterization of its size.

But this is not a satisfactory answer.  While it is certainly plausible that, given some physically salient measure of ``large stress-energy'', large stress-energy will signal the high-energy regime in general relativity, it does not suffice as a definition of that regime because it precludes any vacuum spacetimes---that is, solutions to Einstein's equation with vanishing stress-energy tensor---as having high-energy regions.  This, in turn, means that neither black holes nor high-frequency gravitational waves are necessarily high-energy phenomena; and it would mean that modifications of general relativity that affected only the high-energy regime would not resolve singularities.

There is a natural reply: one needs to consider not (just) stress-energy, but (also) gravitational energy.  This, presumably, is what makes black holes and some gravitational waves high-energy phenomena.  But this answer does not help. ``Gravitational energy'' in general relativity is a deeply vexed problem, with a century of literature highlighting the difficulties.  As \citet{Dewar+Weatherall} argue, the basic problem is that energy is a measure of degree of excitation or deviation from some background standard of motion; but in ``geometrized'' theories such as general relativity, the only physically significant background is the spacetime structure itself.

One can, at least in some cases, introduce further background structure, such as a fixed Minkowski metric at spatial infinity in the case of asymptotically flat spacetimes or a frame field on some region of spacetime, and define energy relative to that.  Some versions of this strategy lead to the definition of so-called energy ``pseudotensors'', which are frame-dependent quantities; others lead to non-local definitions of gravitational energy, such as ADM mass, Bondi energy, or Hawking mass.  Each of these is physically significant in some contexts.  But none of them are suitable for present purposes, for several reasons.  First, whether a theory breaks down should be an invariant fact: it does not depend on a choice of frame field.  And second, as we argued above, we should expect general relativity to break down ``locally''.  Thus \emph{non-local} definitions of energy, such as the ADM mass or Bondi energy, will not suffice, either.\footnote{One might think that \emph{quasi-local} definitions of energy, such as the Hawking mass, would be more satisfactory, since then one could presumably localize energy blow-ups to within certain two-spheres in space-time.  But in fact, this will not work.  Consider, for instance, that the ADM mass of a Schwarzschild black hole is always a finite number $M$; and that a standard desiderata for any quasi-local definition of energy (satisfied by, for instance, the Hawking mass) is that it be both monotonic and asymptote, for appropriate 2-surfaces, to the ADM mass.  It follows that standard quasi-local definitions of energy will be finite (indeed, bounded by $M$) even for regions that ``contain'' the Schwarzschild singularity.}

There is another answer available---though it leaves behind the idea that ``energy'' is the crucial quantity to consider.  In general relativity, it is natural (and common) to associate the ``strong field'' regime with large tidal forces, i.e., large geodesic deviation.  This can be thought of as a measure of the degree to which gravitational influences pull inertial matter at nearby points in different directions.  Tidal forces, meanwhile, are determined by curvature.  One might identify the high-energy regime with the strong-field regime, and thus with large curvature.

Of course, this proposal, too, has challenges.  For one, just as with stress-energy, curvature is a tensor and so it is unclear how to measure its ``size''.  A solution is to construct scalar quantities from curvature, such as the Ricci scalar, $R=R^{ab}{}_{ab}$, or the Kretschman scalar, $K=R^{abcd}R_{abcd}$.  These are physically significant (for instance, the Ricci scalar appears in the Einstein-Hilbert action) and their values can be understood as invariant measures of curvature.  On the other hand, one can generally find examples where only \emph{some} curvature scalars diverge in a given region of spacetime, so it is not clear that any single curvature scalar can suffice as the measure of the high-energy regime.  Indeed: the situation is somewhat worse than this.  There exist singular solutions (e.g. plane waves) of Einstein's equation in which curvature ``becomes large'' but all curvature scalars vanish \citep{Geroch,Ellis+Schmidt}!  These are solutions in which the curvature tensor is null, and thus any contractions yield zero.

Even so, several considerations recommend this proposal.  A preliminary observation is that in other (classical) field theories, such as Maxwell's theory, curvature scalars (i.e., scalars constructed from the field strength $F_{ab}$, which can be interpreted as a curvature tensor) measure energy density.  Moreover, in standard black hole spacetimes, curvature scalars diverge along trajectories approaching a singularity, suggesting that these quantities can capture the sense in which the region near a singularity is where the theory breaks down.  Finally, in effective field theory, it is common to identify the ``high-energy regime'' as one in which terms in an effective Lagrangian become large relative to a fixed energy scale.  But it is the Ricci curvature scalar that appears in the Einstein-Hilbert action (and one would expect other curvature scalars to appear in any extension to that action).  Thus, one might expect the theory to break down when curvature scalars get large relative to an energy cutoff. So the large-curvature regime seems like the mostly likely one for a successor theory to become relevant.

What is the upshot?  We suggest that taking large curvature scalars to signal the high-energy regime is the best option available.  Curvature scalars are defined locally and they are frame/coordinate-independent.  Moreover, they appear in the Einstein-Hilbert action and effective extensions to it and they measure a physically meaningful sense in which ``field strength'' becomes large, corresponding to large tidal forces.  But even so, there is apparently no single, scalar quantity whose large value always and unambiguously signals this regime.  This means we cannot set a scale by a (scalar) cutoff, such that general relativity breaks down (only) when some curvature scalar approaches this value.  Some care is needed in assessing whether a given solution has regions in the ``high-energy'' regime.

\section{Does general relativity fail (only) at high energies?}

Now that we have established what should count as the high-energy regime in general relativity, we can turn to a second question.  To what extent are the ``pathologies'' of general relativity associated with the high-energy regime?  This question is salient because if a higher-energy successor theory should be expected to modify general relativity (only) in the high-energy regime, then such a successor can ``resolve'' these pathologies only if the pathologies themselves occur in the high-energy regime.

Unfortunately, the answer is ``no'', at least on first pass.  Suppose we adopt the common view that singularities signal a failure of general relativity; and suppose we take geodesic incompleteness, i.e., the existence of inextendible geodesics of finite parameter length, as sufficient for a relativistic spacetime to be singular \citep{Hawking+Ellis}.  Then it immediately follows that there are ``low energy'' breakdowns of general relativity.  This is because there exist space-times that are flat, inextendible, and geodesically incomplete \citep{ManchakSHE}.  Such spacetimes are singular but have vanishing curvature (and curvature scalars).  ``High-energy physics'' is not relevant.

How are we to think of such examples?  Philosophers and mathematicians emphasize such examples because they show that there is no necessary connection between, say, curvature and geodesic incompleteness.  On the other hand, one could reasonably argue that such examples are ``unphysical'', and that physically realistic singularities will always be ``curvature singularities'', i.e., that along all incomplete geodesics, some curvature scalar will diverge in finite parameter time.  Curvature singularities are generically the ones that will model singular behavior that could arise via the dynamical co-evolution of matter and geometry, since any kind of dynamical ``collapse'' (or its time reverse, as in expanding FLRW spacetimes) will be associated with unbounded curvature.

Making this sort of intuition fully precise, such that one can identify just that subset of solutions (singular or otherwise) to Einstein's equation that are ``physically reasonable'', is notoriously difficult \citep{ManchakPR}.  But suppose we grant it for present purposes.  In what follows, let us stipulate that only curvature singularities (broadly construed) are ``physical''.  Then it is plausible to think that physically significant singularities \emph{will} be resolved by quantum gravity, since they are associated with the high-energy regime.\footnote{What about other singular spacetimes?  Perhaps those solutions will be eliminated by other means, such as kinematic constraints necessary for quantization.}  Of course, this is a claim about plausibility, since we do not have a theory of quantum gravity in which we can confirm that it holds.  But once again, suppose we grant it.

Does this mean that general relativity breaks down (only) in the high-energy regime, at least once we restrict attention to physically reasonable spacetimes?  It is not clear that it does mean that.  The reason is that singularities are only one of the problematic features of general relativity.  There is another kind of situation where, I would argue, the theory should also be said to break down---or at least, it changes qualitatively in character \citep[c.f.][p. 265]{Hawking+Ellis}.  The feature I have in mind is the existence of extendible maximally globally hyperbolic spacetimes.  The boundary of a maximally globally hyperbolic spacetime across which the spacetime can be extended is known as a ``Cauchy horizon'', and so I will refer to the pathological feature of the theory as the existence of Cauchy horizons.\footnote{Cauchy horizons are often defined more generally, as the boundary of the domain of well-posedness of any initial value problem.  On that definition, Cauchy horizons are not necessarily problematic.  I have in mind specifically the situation where one has a Cauchy horizon in a maximally globally hyperbolic spacetime and the spacetime is extendible.}

To see the problem, suppose one specifies initial data for the vacuum Einstein equation on a surface $S$, satisfying the Einstein constraint equations.  Then there always exists a maximal solution $(M,g_{ab})$, unique up to isometry, which agrees with that initial data under some embedding of $S$ into $M$ \citep{Geroch+Choquet-Bruhat, Hawking+Ellis}.  Any such solution is automatically globally hyperbolic, which means that there exists a Cauchy surface---in this case, the image of $S$ in $M$---which is a surface such that every inextendible smooth timelike curve intersects that surface exactly once.  When this holds, the domain of dependence of $S$, i.e., the collection of points whose field values are determined by the initial data on $S$, is the entire manifold $M$.

There are now several possibilities.  It is possible that $(M,g_{ab})$ is inextendible, i.e., there is no spacetime $(M',g'_{ab})$ such that $(M,g_{ab})$ is isometric to some part of $(M',g'_{ab})$.  In that case, the spacetime that results from evolving the initial data on $S$ is ``as big as it can be''.\footnote{Though see \citet{ManchakExtend} for a discussion of the problems with such modal properties.}  This case is not problematic---or at least, insofar as it \emph{is} problematic, it is because curvature singularities form in the course of the evolution, and we have already discussed that case.  Another possibility is that the spacetime is extendible, but it admits a globally hyperbolic extension.  In that case, evolution could continue in accordance with Einstein's equation, but more data is needed: the surface $S$ has turned out to be too small to uniquely specify what happens beyond $(M,g_{ab})$.  Further data needs to be specified on a Cauchy surface for the extended spacetime.  This case is also unproblematic.

But there is also another possibility.  There exist cases in which, given initial data on a surface $S$, the maximal Cauchy evolution for that initial data admits no globally hyperbolic extension.  Such spacetimes are \emph{maximally globally hyperbolic}.  But the fact that the spacetime is maximally globally hyperbolic does not imply that it does not admit any extension, $(M',g'_{ab})$, at all.  Of course, any extension will not be globally hyperbolic.  But it will nonetheless solve Einstein's equation, everywhere, for some stress-energy tensor $T^{ab}$.  The most famous example of a family of such a spacetimes is the Kerr family, which describe rotating black holes.  In that case, as in many such cases, the reason there is a Cauchy horizon is that the maximal extension of the maximally globally hyperbolic part of Kerr spacetime contains closed timelike curves \citep{ONeill}.  A causal pathology forbids further local evolution but a solution nonetheless exists.

The boundary of the image of $M$ in the extended spacetime manifold $M'$ is the Cauchy horizon for this initial value problem.  It is a surface beyond which the local evolution of the specified initial data cannot continue. Given an extendible maximal globally hyperbolic spacetimes, there is no way to add more initial data to evolve the spacetime further in time.  No amount of information in the past of the Cauchy horizon can uniquely determine how the universe will continue past that horizon---at least, not in the sense of local evolution as determined by Einstein's equation.    Nonetheless, the universe may continue past the horizon.  The result is that the future---i.e., the extension past the Cauchy horizon---is not ``determined'', in the sense of local evolution given by the initial value formulation, by the past.  There is a breakdown of the Einstein evolution equations.

Suppose we stipulate that Cauchy horizons signal a breakdown of general relativity, for the reasons just given or other ones.  Where does this leave us, vis a vis quantum gravity and the high-energy regime?  First, note that Cauchy horizons are physical, insofar as they appear in black hole spacetimes that presumably at least approximately describe real physical situations.  But they are not necessarily high-energy phenomena, in the sense that, like singularities, Cauchy horizons are logically unrelated to curvature.  There exist flat, extendible maximally globally hyperbolic spacetimes, such as Misner spacetime.  And even in Kerr spacetime, curvature is bounded in the vicinity of the Cauchy horizon.

\section{Cauchy horizons, strong cosmic censorship, and the breakdown of GR}

We have just seen that Cauchy horizons, like singularities, are not necessarily a high-energy phenomena.  Is there nonetheless a sense in which the ``physically salient'' Cauchy horizons may be resolved by quantum gravity, just as we argued for singularities?  I suggest that the answer depends on the resolution the strong cosmic censorship hypothesis.  This observation, in turn, suggests a new perspective on the SSCH.  In particular, if the significance of the SSCH is taken to be that it establishes that Cauchy horizons may be (generically) resolved by quantum gravity, then only some versions of the SSCH are physically relevant.

The SSCH may be stated roughly as follows:
\begin{con}
Generically, the maximal Cauchy evolution of (suitable) initial data is (locally) inextendible.
\end{con}
This is not a precise conjecture.  It involves the terms ``generically'' and ``suitable'', which have not be defined. This is a not an idiosyncracy of my presentation; to the contrary, part of the work in settling the conjecture is to identify a precise formulation that captures the underlying idea in a fruitful, physically meaningful way \citep[see][ch. 12]{Wald}.

The SSCH should be interpreted as saying that, except for very special initial conditions, general relativity is deterministic in the sense of local Cauchy evolution.  In other words, Cauchy horizons, of the problematic sort described above, should be extremely rare.  This notion of ``extremely rare'' is supposed to be the converse of ``generic'' in the conjecture; it is sometimes taken to mean that such spacetimes form a sparse set in the space of all spacetimes, relative to some topology.  This in turn means that Cauchy horizons would be unstable under small perturbations, in the sense that if one begins with a spacetime, such as the maximal Cauchy evolution of Kerr initial data, which is maximally globally hyperbolic but also extendible, and one makes arbitrarily small changes to the initial data, the maximal Cauchy evolution of that modified spacetime should be inextendible.  But this gloss does not fully specify what these terms mean, since they are defined only relative to some topology, which we have not specified.

One physical intuition behind the SSCH comes from a famous argument due to \citet{Simpson+Penrose}, based on numerical methods.  The idea is that if a spacetime has a Cauchy surface, signals---in the form, say, of a matter field perturbing the spacetime or gravitational waves---that approach the Cauchy horizon will be blue-shifted to arbitrarily high frequency near the horizon, generating curvature singularities.  These singularities would then prevent the spacetime from continuing beyond the horizon.  More importantly, it would follow that generically Cauchy horizons would fall in the high-energy regime of general relativity.  Curvature scalars would blow up as one approaches the Cauchy horizon.

Many physically interesting formulations of the SSCH are still open.  But recently, \citet{Dafermos+Luk} have settled one precise version.  They show that, assuming the exterior region (i.e., outside of the event horizon) of Kerr spacetimes is stable in a way that they make precise, then for small perturbations to initial data in the interior region of a Kerr black hole, the maximal Cauchy evolution \emph{can be extended} continuously across the Cauchy horizon.  As they put it, ``it will follow that the $C^0$-inextendibility formulation of Penrose's celebrated strong cosmic censorship conjecture is in fact false'' (abstract).  If we interpret ``generic'' in terms of instability in a particular topology (here, a Sobolev topology); and we take ``extendible'' to mean ``continuously (but not necessarily differentiably) extendible'', the SSCH fails.

This work is undoubtedly important progress on a very significant open question.  But given the foregoing, the version of the SSCH that Dafermos and Luk disprove is not the physically salient version.  That is because the existence of $C^0$ extensions of Kerr data across the Cauchy horizon is compatible with the Cauchy horizon being in the ``high-energy'' regime, since it may still be the case that \emph{curvature} blows up as one approaches the horizon.  In other words, even though the metric may be extended across the horizon, it may fail to be (second) differentiable at the horizon.  If so, one should still expect curvature singularities to form as one approaches the Cauchy horizon in Kerr spacetime, and thus for the Cauchy horizon to lie in the high-energy regime.

These remarks suggest that the SCCH should be seen as the conjecture that Cauchy horizons are high-energy phenomena, because they are generically associated with unbounded or undefinable curvature.  From this perspective, falsifying the $C^0$ conjecture does not address the physically relevant question, because only the $C^k$ conjecture, for $k\geq 2$, implies finite curvature extensions.  \citet{Dafermos+Luk} do not settle this version of the conjecture.  But they do make some highly suggestive remarks.  In particular, they argue that in their proof, there are hints that curvature \emph{does} diverge for small perturbations.  They conjecture (but do not prove) that in Kerr spacetime, the maximal Cauchy evolution of perturbed initial data will be $C^2$-inextendible.\footnote{In recent conversations, Luk has expressed continued optimism that this conjecture is true, and indicated he and his collaborators were close to settling it.}

Of course, even if interior Kerr initial data is locally $C^2$-inextendible, it does not follow that the $C^2$-extendibility version of the SSCH is true, even for the topology that Dafermos and Luk adopt.  This is because their conjecture applies only to a single family of examples, whereas the SSCH concerns generic initial data.  But even so, it would provide compelling evidence for what would be, from the perspective of this paper, the most physically relevant SCCH.  In particular, it would show that a version of Penrose's argument would work in one case, suggesting it is likely to work in other cases.  If this is true, then if quantum gravity resolves curvature singularities, one should expect quantum gravity to resolve generic Cauchy horizons as well.

\section{Conclusion}

This paper has addressed two related questions, both salient in the context of the idea that a successor theory to general relativity that modifies that theory in the high-energy regime will ``resolve'' pathologies of the theory, such as singularities. These were, ``what is the high-energy regime of general relativity?''  And ``are all pathologies of general relativity high-energy phenomena?''

In response to the first question, I argued that unbounded curvature should be taken to signal the ``high-energy'' regime of general relativity---even though curvature cannot generally be identified with energy.  I also observed, however, that measuring ``large curvature'' is subtle and that in general, there is no single scalar quantity whose divergence can be identified with the high-energy regime. Instead, one must consider both various curvature scalars and the behavior of the Riemann tensor itself to assess whether a given region of a spacetime is ``high-energy''.  One cannot identify a fixed energy cut-off and claim that general relativity breaks down whenever a single scalar quantity exceeds that value. 

In response to the second question, I argued that strictly speaking the answer is ``no''.  For instance, singularities can occur even in flat spacetime, and thus not all singularities are associated with divergent curvature.  On the other hand, I also argued that the most physically salient singularities are curvature singularities, which are singularities with the property that curvature scalars diverge as one approaches the singularity.  These are the singularities that can emerge dynamically, such as through gravitational collapse or via divergent matter fields.  And these singularities \emph{are} associated, by definition, with large curvature, and thus may be expected to be resolved by a high-energy successor to general relativity.

Finally, I argued that whether another class of pathologies in general relativity---namely, the Cauchy horizons in extendible maximally globally hyperbolic spacetimes---should be associated with the high-energy regime in general relativity depends on the resolution of a particular version of the SSCH.  This version of the SCCH would forbid the $C^2$ extendibility of the maximal Cauchy evolution of generic initial data.  I noted that this is a weaker conjecture than the $C^0$ version recently shown to be false by \citet{Dafermos+Luk}; and I argued that if a conjecture due to Dafermos and Luk, that perturbed Kerr initial data is $C^2$ inextendible, it would provide evidence for this version of the SSCH.  The upshot was a new interpretation of the physical significance of the SSCH, which is that if the $C^2$-extendibility version of the hypothesis holds, then the sort of breakdown of general relativity signaled by failures of the local evolution equation may be resolved by a high-energy successor theory.

\section*{Acknowledgments} This project is based on work supported by the John Templeton Foundation grant “New Directions in Philosophy of Cosmology” (grant no. 61048). I am grateful to Feraz Azhar, Jeremy Butterfield, Erik Curiel, Juliusz Doboszewski, Sam Fletcher, Henrique Gomes, Sean Gryb, Eleanor Knox, Serge Rudaz, Bob Wald for very helpful discussions in connection with this work.  The paper benefited greatly from feedback from audiences at the Harvard Black Hole Initiative, Cambridge University, the conference Logic, Relativity, and Beyond 20/21, and the 20th UK/European Foundations of Physics meeting in Paris.

\bibliography{bh}
\bibliographystyle{elsarticle-harv}

\end{document}